\documentclass[10pt,aps,prd,twocolumn,showpacs,superscriptaddress,nofootinbib,nobibnotes,longbibliography,floatfix]{revtex4-2}

\usepackage[utf8]{inputenc}
\usepackage[T1]{fontenc}
\usepackage{bm}
\usepackage{mathtools,amsmath,amssymb,amsfonts,mathrsfs,eucal,graphicx,tensor,csquotes,accents,commath,chngcntr,siunitx}
\usepackage[dvipsnames]{xcolor}
\usepackage[unicode]{hyperref}
\hypersetup{colorlinks=true, citecolor=MidnightBlue,
            linkcolor=MidnightBlue, urlcolor=MidnightBlue, linktocpage=true}
\usepackage[normalem]{ulem}
\usepackage{orcidlink}

\newcommand{\orcid}[1]{\href{https://orcid.org/#1}{\includegraphics[width=10pt]{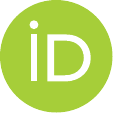}}}

\begin{document}

\author{Saulo Albuquerque \orcid{0000-0003-2911-9358}}
\email{saulo.soaresdealbuquerquefilho@uniurb.it}
\affiliation{Theoretical Astrophysics, IAAT, University of T\"ubingen, Auf der Morgenstelle 10, D-72076 T\"ubingen, Germany}

\author{Sebastian H.\,V\"olkel \orcid{0000-0002-9432-7690}}
\email{sebastian.voelkel@uni-tuebingen.de}
\affiliation{Theoretical Astrophysics, IAAT, University of T\"ubingen, Auf der Morgenstelle 10, D-72076 T\"ubingen, Germany}

\date{\today}

\title{Optimal frequency scales for probing black-hole geometries}

\begin{abstract}
Can gravitational waves probe the near-horizon geometry of black holes, and if yes, which frequency scale is optimal? Although shorter wavelengths usually resolve smaller scales, we show that black-hole scattering may impose an information-theoretic optimum. We study a controlled scattering Gedankenexperiment in which Gaussian pulses of scalar test fields are sent toward a black-hole potential and the reflected waveform is used to infer the geometry. Near-horizon deviations are parametrized with the Rezzolla-Zhidenko metric, and information recovery is quantified by the Fisher matrix in the high-signal-to-noise limit. Narrow, high-frequency pulses resolve short scales but are mostly transmitted through the barrier, while wide pulses are efficiently reflected but poorly resolve the potential. Their competition selects an optimal pulse width, numerically found to be set by about the inverse square root of the potential peak. Using the P\"oschl-Teller analytical solutions, we further model the correct excitation of quasinormal modes and separate the information in the fundamental mode from that in the full waveform, including the prompt response. The optimal probe is therefore not the highest-frequency pulse, but the waveform that balances spatial resolution against reflected information, linking black-hole spectroscopy, semiclassical barrier scattering, and information theory. 
\end{abstract}

\maketitle

\section{Introduction}
Controlled scattering experiments have been central to modern physics. Rutherford scattering revealed the compact atomic nucleus~\cite{Rutherford:1911zz}, and particle colliders now use the same basic idea to probe the constituents and interactions of the Standard Model~\cite{ParticleDataGroup:2024cfk}. 
For black holes and gravitational waves, no similar, controlled experiment is available. Instead, information must be inferred from astrophysical systems selected by nature rather than prepared in the laboratory~\cite{LIGOScientific:2016aoc}. 
Thus, only sufficiently violent events occurring within an observational timescale can be observed, while the same system cannot be tuned or repeatedly probed at will. 
This limits both the range of accessible sources and the degree of experimental control. 

A key question is whether the near-horizon geometry of astrophysical black holes leaves observable traces in gravitational radiation, e.g., in quasinormal modes~\cite{Kokkotas:1999bd,Berti:2025hly}. 
Most work on black-hole spectroscopy has focused on quasinormal modes after merger, whose frequencies are largely controlled by the effective potential near the light ring rather than by the horizon itself~\cite{Schutz:1985km}. 
Although overtones are more sensitive to the near-horizon geometry~\cite{Konoplya:2022pbc}, they are also more difficult to extract~\cite{Nee:2023osy,Baibhav:2023clw}. 
Direct waves have been discussed as another possible probe of the near-horizon structure~\cite{Mino:2008at,Zimmerman:2011dx,oshita2025probingdirectwavesblack,Lu:2025vol}. 
It is therefore important to understand, in a controlled setting, how near-horizon information can be encoded in scattered waves. 
A large body of literature has studied modifications of the near-horizon geometry and boundary conditions close to the would-be horizon of exotic compact objects, for which echoes can provide a characteristic signature~\cite{Kokkotas:1995av,Ferrari:2000sr,Tominaga:1999iy,Cardoso:2016rao}. 
Here, we instead consider black holes with an absorbing horizon and ask how well their near-horizon geometry can be constrained. 

To explore this question, we extend Vishveshwara's pioneering work~\cite{Vishveshwara:1970zz}, and consider a controlled scattering Gedankenexperiment in which an observer sends a wave pulse toward a black hole and infers the spacetime geometry from the reflected signal. We adopt the Rezzolla-Zhidenko (RZ) parametrization of nonrotating black holes~\cite{Rezzolla:2014mua}, evolve a massless scalar field in the time domain, and use the Fisher information matrix~\cite{Finn:1992wt,Cutler:1994ys} to study how the properties of the incident pulse affect the reconstruction of the black-hole metric and effective potential in the high signal-to-noise-ratio (SNR) limit. 

We find that the pulse width determines a competition between spatial localization and reflectivity. Narrow pulses contain more high-frequency power and can more easily penetrate the potential barrier, thereby accessing structure closer to the horizon, but a larger fraction of the signal is lost into the black hole. Wider pulses are reflected more efficiently, but their predominantly low-frequency content carries less information about the near-horizon geometry. This trade-off is balanced by a well-defined optimal pulse width for parameter inference. For the massless scalar-field wave equation, we show numerically that the optimum is set by the height of the effective-potential barrier, $\sigma_{\rm opt}\sim 1/\sqrt{V_{\rm max}}$, and reproduce this scaling analytically for the P\"oschl-Teller (PT) potential. Importantly, the optimal scale is similar across RZ parameters of different orders, even though higher-order parameters encode increasingly near-horizon structure. Thus, higher-frequency waves do not necessarily provide stronger constraints on smaller scales close to the horizon, as one might naively expect from standard problems in quantum mechanics. 

Related optimization questions were previously studied in Ref.~\cite{Doukas:2014xja} using quasimonochromatic scalar scattering from the Regge-Wheeler potential and its P\"oschl–Teller approximation~\cite{Poschl:1933zz,Ferrari:1984ozr,Ferrari:1984zz,Kuntz:2025gdq} to identify an optimal frequency close to the one reported here. 
In contrast, we consider finite Gaussian wave packets and multiparameter RZ geometries in a time-domain scattering framework, using the P\"oschl-Teller potential to compare the information encoded in the complete scattering response with that contained in the excited fundamental quasinormal mode. 
More specifically, we quantify the information gain obtained by analyzing both, the full scattering response, and the excited fundamental quasinormal mode. 
We find that the two contain comparable information at high frequencies and near the optimal scale, whereas the fundamental quasinormal mode alone becomes less informative at lower frequencies. The validity of the fundamental quasinormal approximation near the optimal scale allows us to analytically demonstrate the aforementioned scaling law. Throughout this work, we use units in which $G=c=1$.

\section{Methods}\label{methods}

\emph{\bf Scattering experiment--} 
As a starting point for a prototypical black hole scattering setup, we consider the massless scalar wave equation
\begin{equation}
\Box \Phi
\equiv
\frac{1}{\sqrt{-g}}
\partial_\mu
\left(
\sqrt{-g}\,g^{\mu\nu}\partial_\nu \Phi
\right)
=0\,,
\label{eq_wave}
\end{equation}
on a static, spherically symmetric black-hole background,
\begin{equation}
ds^2=-N^2(r)dt^2+\frac{B^2(r)}{N^2(r)}dr^2+r^2d\Omega^2\,.
\label{eq:metric}
\end{equation}
Since actual gravitational waves require a gravitational theory for prescribing their dynamics, the analysis would either be theory-specific, or involve a more complicated test-field calculation. 
Because scalar-field perturbations cover many aspects of the full gravitational wave case in general relativity, we consider it as a clean setup for an exploration of this kind. 

To cover a wide range of black hole geometries, we describe the background using the RZ parametrization~\cite{Rezzolla:2014mua}. 
It introduces the compact coordinate $x=1-r_0/r$, where $r_0$ is the horizon radius, and expands the metric functions $N(r)$ and $B(r)$ in powers of $1-x$ and continued fractions. 
This separates the asymptotic behavior from corrections that become increasingly important near the horizon and provides a rapidly convergent description of the full exterior spacetime~\cite{Konoplya:2020hyk}. 
The RZ parameters $\bm{\theta}$ entering at different orders of the expansion are treated as free parameters in the inference problem. 
Further details are given in Appendix~\ref{app_RZ} and in the original work~\cite{Rezzolla:2014mua}. 

Decomposing the scalar field into spherical harmonics reduces Eq.~\eqref{eq_wave} to the one-dimensional wave equation
\begin{equation}
\left[
\partial_t^2-\partial_{r_*}^2+V(r)
\right]\Psi(t,r_*)=0\,,
\label{eq:RW}
\end{equation}
where the tortoise coordinate is defined by $dr_*/dr=B/N^2$. 
The effective potential is determined by the background metric and is given by
\begin{equation}
V(r)=N^2(r)\left[
\frac{\ell(\ell+1)}{r^2}
+\frac{1}{rB(r)}
\frac{d}{dr}\left(\frac{N^2(r)}{B(r)}\right)
\right]\,.
\label{eq:potential}
\end{equation}

The initial data are chosen as an ingoing Gaussian pulse,
\begin{equation}
\Psi(0,r_*)=
A\exp{\left[-\frac{(r^*-r^*_c)^2}{\sigma^2}\right]}\,,
\label{eq:gaussian}
\end{equation}
centered at $r^*_c$ well outside the potential barrier. 
Its Fourier amplitude scales as $\tilde{\Psi}(\omega)\propto \sigma\exp(-\sigma^2\omega^2/4)$, so that narrow pulses contain a broader range of frequencies, whereas wider pulses are dominated by low-frequency components. 
After interacting with the potential, part of the wave is transmitted toward the horizon and part is reflected back to the observer. 
We record the reflected waveform at large distances and use it to infer the RZ parameters. 
The numerical time-domain integration using finite differences 
is discussed in Appendix~\ref{app_TD}.

\medskip
\emph{\bf Data analysis setup--} 
We denote the reflected waveform measured by the observer by \(h(t,\bm{\theta})\), where \(\bm{\theta}\) are the RZ parameters varied in the analysis. 
Assuming stationary Gaussian noise, the likelihood for observed data \(d(t)\) is proportional to
\begin{equation}
p(d|\bm{\theta})
\propto
\exp\left[
-\frac{1}{2}
\left(
d-h(\bm{\theta})
\middle|
d-h(\bm{\theta})
\right)
\right]\,.
\label{eq:likelihood}
\end{equation}
For the proof-of-principle analysis considered here, we assume white noise and define the time-domain inner product as
\begin{equation}
(a|b)
=
\frac{1}{S}
\int_{t_{\rm start}}^{t_{\rm end}}
a(t)b(t) {\rm d}t,
\label{eq:inner_product}
\end{equation}
where \(S\) is a constant noise level. 
Note that a frequency-dependent sensitivity would of course impact data analysis results, but this is purely a detector-related effect, and not fundamental to the wave scattering problem. 
The integration interval is chosen to contain the complete reflected response, including the prompt response, ringdown, and power-law tails at late times.

In the high-SNR limit, the likelihood around the considered parameters is approximately Gaussian and the covariance matrix $\Sigma_{ij}$ can be determined from the Fisher-information matrix,
\begin{equation}
\Gamma_{ij}
=
\left(
\partial_i h
\middle|
\partial_j h
\right),
\qquad
\Sigma_{ij}
\simeq
\left(\Gamma^{-1}\right)_{ij}\,,
\label{eq_fisher}
\end{equation}
and the marginalized uncertainty of a parameter \(\theta_i\) via
\begin{equation}
\Delta\theta_i
=
\sqrt{\Sigma _{ii}}\,.
\label{eq:parameter_error}
\end{equation}

To isolate the effect of the spectral content of the initial pulse, we vary its width \(\sigma\) while keeping its SNR fixed,
\begin{equation}
\rho_{\rm in}^{2}
=
\left(
h_{\rm in}
\middle|
h_{\rm in}
\right)\,,
\label{eq:incident_snr}
\end{equation}
where $h_{\rm in}$ is the initial Gaussian pulse sent by the observer to probe the black hole. 
For each value of \(\sigma\), the incident amplitude is rescaled so that \(\rho_{\rm in}\) remains unchanged. 
This normalization removes the trivial dependency of the parameter accuracy by increasing/decreasing the SNR of the incoming pulse. 
Note that the value of $S$ in the high-SNR limit is only a scaling factor. 
The Fisher matrix is then obtained for the reflected waveform $h_{\rm out}$, quantifying the observer's capability to probe the black hole geometry. 

\section{Results}\label{results}

\emph{\bf Inferring RZ metric--} 
We start our discussion by examining the measurement uncertainties, $\Delta\theta_i$, associated with RZ parameters of different orders within the Fisher-matrix approximation of the Gaussian wave packet scattering experiment. 

\begin{figure}
\centering
\includegraphics[width=1.0\linewidth]{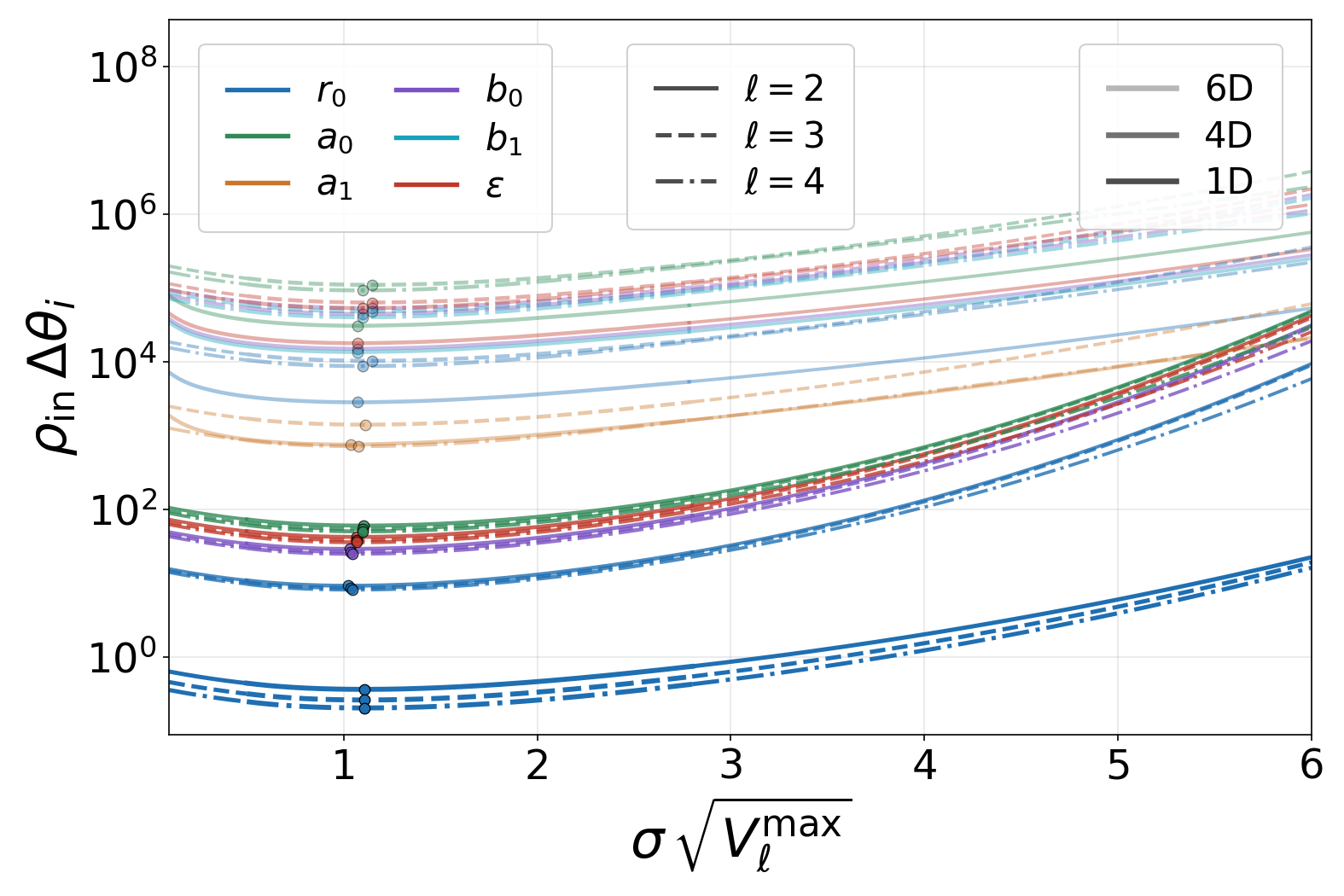}
\caption{We show the measurement errors $\Delta \theta_i$ of various RZ parameters as a function of the wave packet width $\sigma$ times $\sqrt{V_{\ell}^{\mathrm{max}}}$. 
Different colors represent different RZ parameters, while different linestyles distinguish results from independent angular numbers $\ell$, and opacity highlights the dimensionality of the covariance matrix, i.e., the number of simultaneously varied parameters, where 1D:$(r_0)$, 4D:$(r_0, a_0, b_0, \epsilon)$, and 6D:$(r_0, a_0, a_1, b_0, b_1, \epsilon)$. \label{scaling}}
\end{figure}

Figure~\ref{scaling} shows the marginalized measurement uncertainties as functions of the dimensionless combination $\sigma\sqrt{V_{\ell}^{\mathrm{max}}}$, where $\sigma$ denotes the wave-packet width and $V_{\ell}^{\mathrm{max}}$ is the maximum of the effective potential for a given angular multipole number $\ell$. 
To explore a broad range of scenarios, we vary both $\ell$ and the total number of RZ parameters included simultaneously in the analysis. 
Remarkably, the marginalized uncertainties display a clear universal behavior across all RZ parameters, multipole numbers, and parameter-set sizes: in each case, there is an optimal wave-packet width that minimizes the uncertainty.
This optimal scale is approximately
\begin{align}
\sigma_{\mathrm{optimal}} \sim \frac{1}{\sqrt{V_{\ell}^{\mathrm{max}}}}\,,
\end{align}
and corresponds to the characteristic inverse-frequency scale associated with the real part of the fundamental quasinormal mode frequency, as estimated using the Schutz-Will formula~\cite{Schutz:1985km}, and in agreement with Ref.~\cite{Doukas:2014xja}. 
Note that, although $V^\mathrm{max}_{\ell}$ scales proportional to $\ell(\ell+1)$ and therefore, $\sigma_{\rm optimal}$ decreases, it  always corresponds to the region around the maximum of the potential. 
Although increasing $\ell$ reduces the measurement errors, the dependency on the number of simultaneously varied parameters is more significant. 

The scaling of $\Delta \theta_i$ for larger $\sigma \gg \sigma_{\mathrm{optimal}}$ should be expected, because low-frequency waves are mostly reflected far away from the black hole, making precision measurements of the near-horizon structure challenging. 
On the other hand, our results for $\sigma \ll \sigma_{\mathrm{optimal}}$ are less trivial, since they are counterintuitive. 
Although the higher-frequency waves can propagate through the potential, there is no mechanism to reflect them back. 
Possible changes in the near-horizon geometry, manifested in large changes in the effective potential behind its standard maximum, would require significant changes in the geometry. 
Note that such modifications are, in principle, possible to describe with the RZ metric~\cite{Konoplya:2022tvv}, but this implies that small changes cannot be constrained well. 

\begin{figure}
\centering
\includegraphics[width=1.0\linewidth]{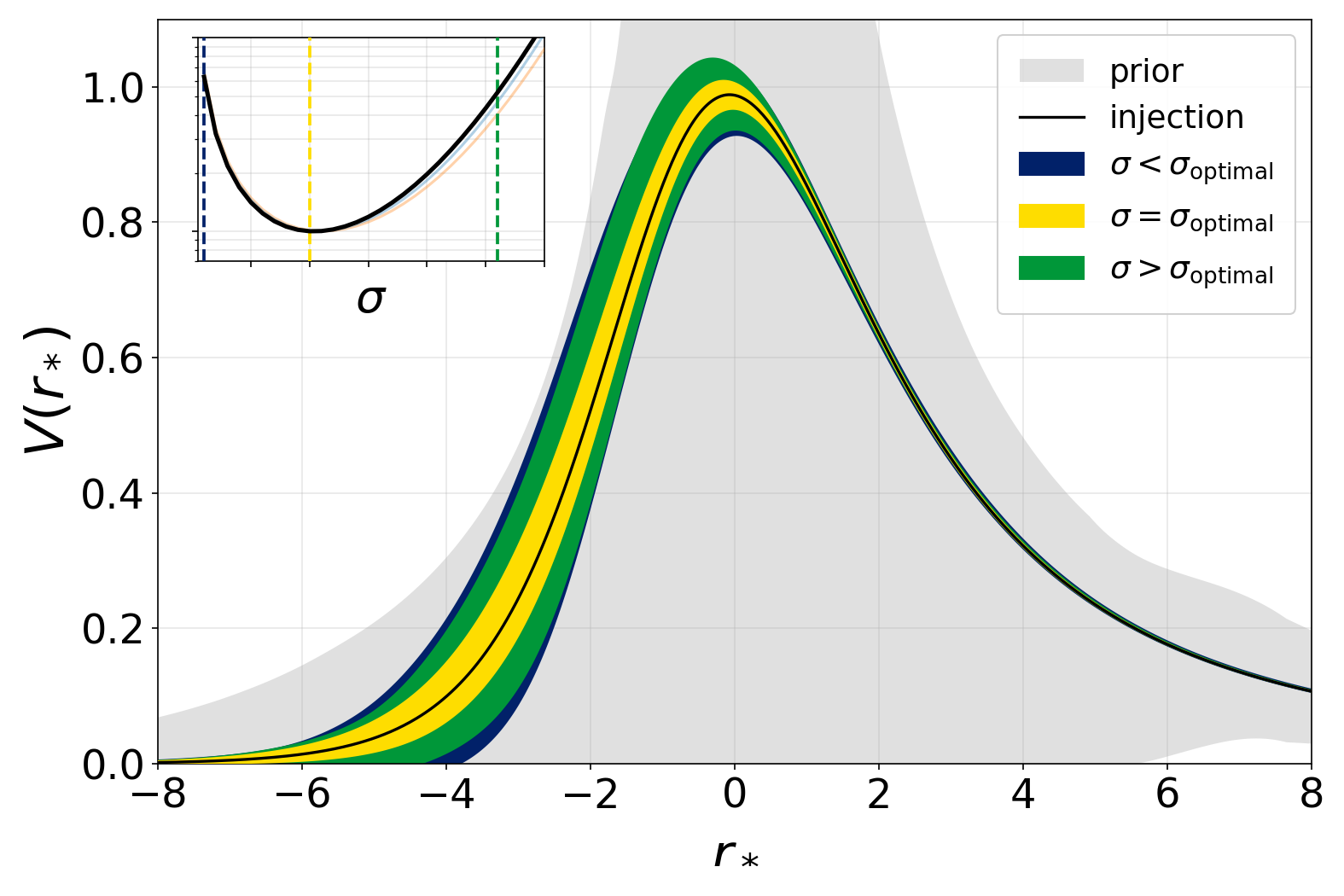}
\caption{We show the injected effective potential (black line) and compare it to $68\,\%$ HDIs computed for three different $\sigma$ indicated in the inset (different colors), and the prior. For generating this plot, $\rho_{\rm in}=250$ was used. 
}
\label{potential}
\end{figure}

\medskip
\emph{\bf Inferring effective potential--} 
To quantify these considerations, we now examine what the constraints on the metric parameters imply for our knowledge of the effective potential, which is probed more directly by wave scattering. 
In Fig.~\ref{potential}, we present posterior constraints in the form of highest-density intervals (HDIs). 
These intervals are obtained by sampling the RZ parameters from the covariance matrices for $\ell=2$ and rejecting samples that yield unphysical black-hole metrics, such as, for example, spacetime realizations where $g_{tt}$ and/or $g_{rr}$  have non-positive values at any point in the radial domain; see Ref.~\cite{Albuquerque:2025eny} for further details and Bayesian analysis.

The resulting uncertainty bands quantify how tightly the effective potential is constrained. 
They show that the outer side of the potential barrier is significantly better constrained, as expected for the scattering setup considered here. 
By contrast, the uncertainty increases toward the inner side of the barrier. 
The apparent vanishing of this uncertainty near the horizon is imposed by the functional form of $g_{tt}$ in the RZ parametrization and is therefore a consequence of the parametrization itself. 
It should not be interpreted as evidence that the effective potential is tightly constrained in the near-horizon region.
To demonstrate that this asymmetry is not primarily driven by the larger metric uncertainty near the horizon, we also show the HDI corresponding to the RZ priors, revealing that both sides of the potential are flexible.  

\medskip
\emph{\bf Analytical results--}
The observation of an optimal probing scale for different $\ell$ and different number of RZ parameters suggests that the minima observed in Fig.~\ref{scaling} are not merely a numerical feature of the RZ parametrization or Regge-Wheeler potential structure, but instead reflects a more general property of wave scattering. To prove this fact, we analytically demonstrate the existence of an optimal Gaussian packet width for the widely-used P\"oschl-Teller potential~\cite{Ferrari:1984ozr,Ferrari:1984zz}, which provides an accurate approximation of the peak of the angular momentum barrier in Eq.\eqref{eq:potential}, while admitting closed-form solutions. 
It is given by
\begin{equation}
V(r^*)
=
\frac{V_0}
{\cosh^2\!\left[\alpha(r^*-r^{*}_0)\right]}\,,
\label{eq:PT}
\end{equation}
where $V_0$ is the barrier height, $\alpha$ determines its curvature, and $r_{0}^*$ denotes the position of the maximum. 
The corresponding quasinormal-mode spectrum, waveform solutions, and scattering coefficients are known analytically~\cite{Ferrari:1984ozr,Ferrari:1984zz}. 

Hence, the derivation of the Fisher matrix for the free parameters $\bm{\theta}^{\mathrm{PT}}=\left( r^*_0, V_0, \alpha \right)$ of the P\"oschl-Teller potential can, in practice, be carried out analytically. 
Further details are discussed in Appendix \ref{app_PT}, where we also investigate an approximation formula valid near the dominant resonance (the fundamental quasinormal mode), namely
\begin{equation}
\Sigma_{ij}=\left(\Gamma^{-1}\right)_{ij}
\propto
\frac{1}{\sigma}
\exp
\left[
\frac{\sigma^2\omega_R^2}{2}
\right]\,,
\label{eq:covanalytic}
\end{equation}
where  $\omega_R$ is the oscillation frequency of the fundamental quasinormal mode. 
In addition to the extra functions dependent on $\sigma$ in this equation, an approximately general relation between $\sigma_{\rm opt}$ and $V_0$ can be established , namely 
\begin{equation}
\sigma_{\rm opt}
\propto \frac{1}{\omega_R}=
\frac{1}
{\sqrt{V_0}\sqrt{1-\alpha^2/(4V_0)}}
\simeq
\frac1{\sqrt{V_0}}\,,
\label{eq:finalopt}
\end{equation}
where the final approximation is valid whenever $\alpha^2\ll4V_0$, which is the case when matching to the Regge-Wheeler potential. 

\begin{figure}
\centering
\includegraphics[width=1.0\linewidth]{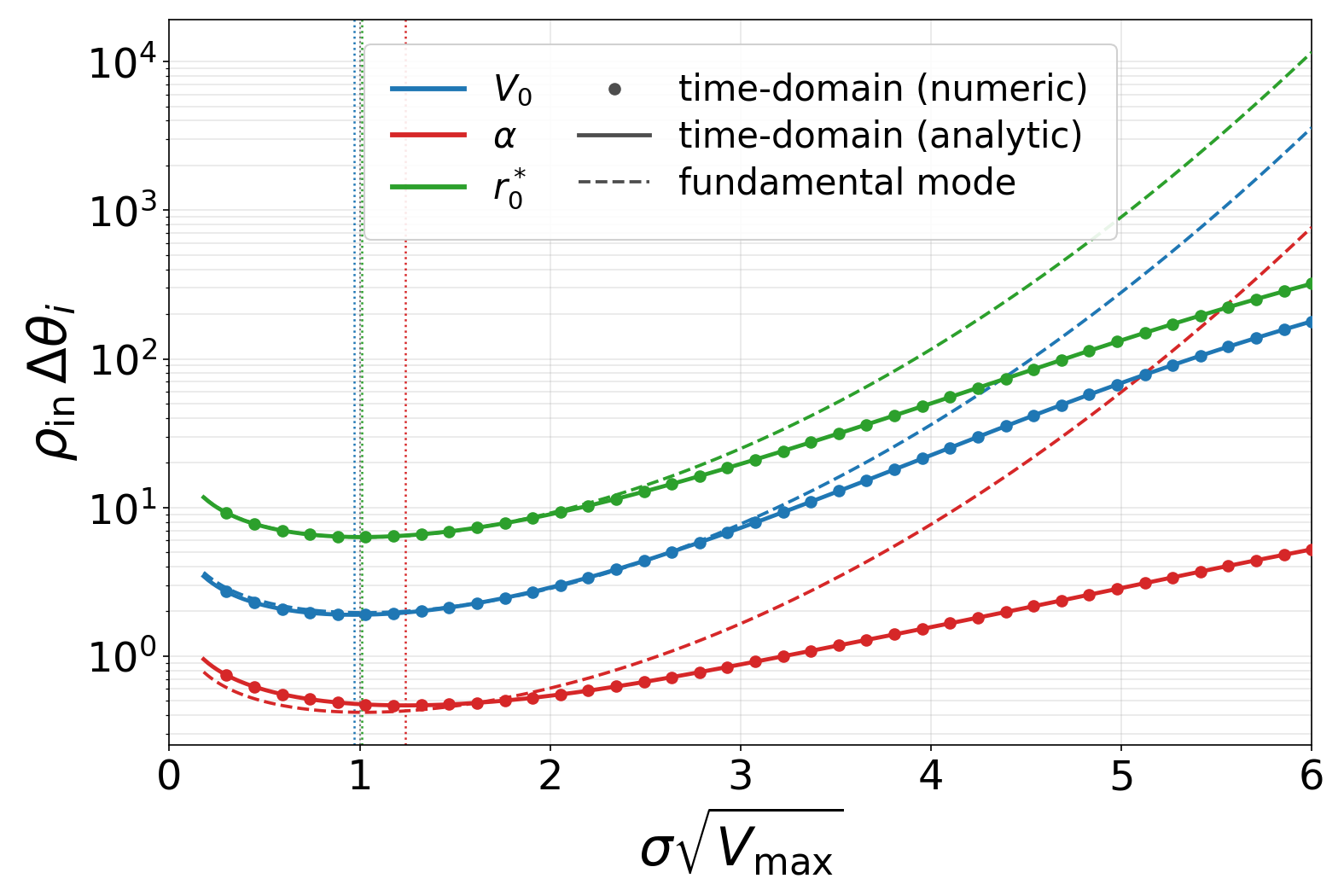}
\caption{We show the measurement errors $\Delta \theta_i$ times the ingoing wave packet SNR $\rho_{\mathrm{in}}$ as a function of the wave packet width $\sigma$ times $\sqrt{V_{\mathrm{max}}}$. 
Different parameter of the P\"oschl-Teller potential are shown with different colors and linestyles represent different ways of obtaining them. 
}\label{fig_ptcase}
\end{figure}

In Fig.~\ref{fig_ptcase}, we compare the fundamental mode approximation Eq.~\eqref{eq:covanalytic} with the full analytical prediction of the Fisher matrix (which we invert numerically), and the numerical time-domain Fisher analysis (as for the RZ case previously). 
The agreement between the analytical and numerical Fisher analysis is excellent, and can serve as an independent check for the numerical accuracy of our time-domain code.

The most important aspect is the information in the fundamental mode approximation. 
For scales $\sigma \lesssim \sigma_{\mathrm{optimal}}$, it is in very good agreement, and provides roughly similar results for the location of the minimum error (with small differences for $\alpha$). 
However, for $\sigma \gg \sigma_{\mathrm{optimal}}$ there is a systematic trend for even larger measurement uncertainties compared to the full analysis. 
This is reasonable, as large $\sigma$ corresponds to low-frequency waves, which are mostly reflected before exciting the fundamental mode around the maximum of the potential, and complements the qualitative discussion for the RZ parameters. 

This part of our analysis demonstrates, to our knowledge for the first time, that the Fisher information in such scattering experiments is closely related to fundamental-mode excitation, even for high-frequency wave packets, whereas this identification fails for low-frequency waves.

\section{Conclusions}\label{conclusions}

We conducted an idealized black-hole scattering experiment to quantify the question whether scattered waves probe the near-horizon geometry of black holes, and if yes, which frequency scale is optimal. 
Our results suggest that scattering experiments select a preferred frequency scale for recovering information about the underlying geometry. 
Although narrow pulses provide greater spatial resolution and penetrate further toward the horizon, much of their power is transmitted through the effective-potential barrier and is therefore lost to the observer. 
Wider pulses are reflected more efficiently, but their predominantly low-frequency content is less sensitive to the structure of the potential, and thus the near-horizon black hole geometry. 
The competition between these effects yields an optimal pulse width for parameter inference. 

For massless scalar perturbations in the RZ spacetime, we first demonstrated numerically that the optimal width of a Gaussian wave packet with fixed initial SNR is well-approximated by $\sigma_{\rm optimal}\sim 1/\sqrt{V^{\max}_{\ell}}$. 
Remarkably, the optimal scale remains similar for RZ parameters entering at different orders, even though higher-order coefficients encode increasingly near-horizon modifications. 
Thus, probing smaller scales near the horizon does not generally favor arbitrarily high-frequency waves. 
Instead, the optimal probe is determined by a balance between spatial resolution and the amount of information reflected back to the observer. 

To quantify our findings further, we then recovered the same behavior analytically by using the widely-used P\"oschl-Teller potential as an approximate model. 
In this case, we also quantified how much information is contained in the excited fundamental mode, and how much in the full signal. 
This revealed that the identification of the fundamental mode is reliable for intermediate- and high-frequency wave packets, but fails for low-frequency packets. 
It remains an intriguing open question to explore similar scattering experiments for rotating black holes, as new phenomena may arise that could potentially generalize some of our findings. 

While the studied framework is a pure Gedankenexperiment, analog gravity systems~\cite{Barcelo:2005fc} may provide the opportunity to explore very similar scattering problems in actual laboratory experiments. 
Such platforms have already enabled controlled studies of curved-spacetime phenomena, from analogue Hawking radiation and rotational superradiance to quasinormal-mode oscillations and black-hole spectroscopy~\cite{Unruh:1980cg,Steinhauer:2015saa,Torres:2016iee,Torres:2020tzs,smaniotto2026blackholespectroscopygiantquantum}. 
Bayesian parameter estimation in related analog-gravity settings has recently been considered in Ref.~\cite{Albuquerque:2025eny}, while Ref.~\cite{solidoro2026spectroscopyanalogueblackholes} introduced simulation-based inference to account for the intrinsic stochasticity conjectured in such experiments~\cite{Solidoro:2024yxi}. 
Our results suggest that such systems could also test how the frequency content of an incident perturbation controls the geometric information recovered from the reflected signal. 

\smallskip
\acknowledgments 
S.\,A acknowledges support from the Alexander von Humboldt Foundation. 
This study was financed in part by the Coordenação de Aperfeiçoamento de Pessoal de Nível Superior – Brasil (CAPES) – Finance Code 001.

\bibliography{literature}

@article{Lu:2025vol,
    author = "Lu, Neil and Ma, Sizheng and Piccinni, Ornella J. and Chen, Yanbei and Sun, Ling",
    title = "{GW250114 reveals signatures of post-merger black-hole horizon}",
    eprint = "2510.01001",
    archivePrefix = "arXiv",
    primaryClass = "gr-qc",
    doi = "10.1038/s41586-026-10696-0",
    journal = "Nature",
    volume = "655",
    number = "8121",
    pages = "71--75",
    year = "2026"
}

@article{Konoplya:2022pbc,
    author = "Konoplya, R. A. and Zhidenko, A.",
    title = "{First few overtones probe the event horizon geometry}",
    eprint = "2209.00679",
    archivePrefix = "arXiv",
    primaryClass = "gr-qc",
    doi = "10.1016/j.jheap.2024.10.015",
    journal = "JHEAp",
    volume = "44",
    pages = "419--426",
    year = "2024"
}

@article{Berti:2025hly,
    author = "Berti, Emanuele and others",
    editor = "Berti, Emanuele and Cardoso, Vitor and Carullo, Gregorio",
    title = "{Black hole spectroscopy: from theory to experiment}",
    eprint = "2505.23895",
    archivePrefix = "arXiv",
    primaryClass = "gr-qc",
    doi = "10.1088/1361-6382/ae59e2",
    journal = "Class. Quant. Grav.",
    volume = "43",
    number = "12",
    pages = "123001",
    year = "2026"
}

@article{Albuquerque:2025eny,
    author = {Albuquerque, Saulo and V{\"o}lkel, Sebastian H.},
    title = "{Bayesian analysis of analog gravity systems with the Rezzolla-Zhidenko metric}",
    eprint = "2501.09000",
    archivePrefix = "arXiv",
    primaryClass = "gr-qc",
    doi = "10.1103/kwrg-rs71",
    journal = "Phys. Rev. D",
    volume = "111",
    number = "12",
    pages = "124020",
    year = "2025"
}

@article{Doukas:2014xja,
    author = "Doukas, Jason and Westwood, Luke and Faccio, Daniele and Di Falco, Andrea and Fuentes, Ivette",
    title = "{Gravitational parameter estimation in a waveguide}",
    eprint = "1403.4855",
    archivePrefix = "arXiv",
    primaryClass = "quant-ph",
    doi = "10.1103/PhysRevD.90.024022",
    journal = "Phys. Rev. D",
    volume = "90",
    number = "2",
    pages = "024022",
    year = "2014"
}

@article{Barcelo:2005fc,
    author = "Barcelo, Carlos and Liberati, Stefano and Visser, Matt",
    title = "{Analogue gravity}",
    eprint = "gr-qc/0505065",
    archivePrefix = "arXiv",
    doi = "10.12942/lrr-2005-12",
    journal = "Living Rev. Rel.",
    volume = "8",
    pages = "12",
    year = "2005"
}

@misc{solidoro2026spectroscopyanalogueblackholes,
      title={Spectroscopy of analogue black holes using simulation-based inference}, 
      author={Leonardo Solidoro and Sebastian H. Völkel and Silke Weinfurtner},
      year={2026},
      eprint={2604.12800},
      archivePrefix={arXiv},
      primaryClass={gr-qc},
      url={https://arxiv.org/abs/2604.12800}, 
}

@article{Rezzolla:2014mua,
    author = "Rezzolla, Luciano and Zhidenko, Alexander",
    title = "{New parametrization for spherically symmetric black holes in metric theories of gravity}",
    eprint = "1407.3086",
    archivePrefix = "arXiv",
    primaryClass = "gr-qc",
    doi = "10.1103/PhysRevD.90.084009",
    journal = "Phys. Rev. D",
    volume = "90",
    number = "8",
    pages = "084009",
    year = "2014"
}

@article{Schutz:1985km,
    author = "Schutz, Bernard F. and Will, Clifford M.",
    title = "{Black hole normal modes: A semianalytic approach}",
    reportNumber = "PRINT-85-0063 (WASH.U.,ST.LOUIS)",
    doi = "10.1086/184453",
    journal = "Astrophys. J. Lett.",
    volume = "291",
    pages = "L33--L36",
    year = "1985"
}

@article{Cardoso:2016rao,
    author = "Cardoso, Vitor and Franzin, Edgardo and Pani, Paolo",
    title = "{Is the gravitational-wave ringdown a probe of the event horizon?}",
    eprint = "1602.07309",
    archivePrefix = "arXiv",
    primaryClass = "gr-qc",
    doi = "10.1103/PhysRevLett.116.171101",
    journal = "Phys. Rev. Lett.",
    volume = "116",
    number = "17",
    pages = "171101",
    year = "2016",
    note = "[Erratum: Phys.Rev.Lett. 117, 089902 (2016)]"
}

@inproceedings{Kokkotas:1995av,
    author = "Kokkotas, K. D.",
    title = "{Pulsating relativistic stars}",
    booktitle = "{Les Houches School of Physics: Astrophysical Sources of Gravitational Radiation}",
    eprint = "gr-qc/9603024",
    archivePrefix = "arXiv",
    pages = "89--102",
    month = "9",
    year = "1995"
}

@article{Ferrari:2000sr,
    author = "Ferrari, V. and Kokkotas, K. D.",
    title = "{Scattering of particles by neutron stars: Time evolutions for axial perturbations}",
    eprint = "gr-qc/0008057",
    archivePrefix = "arXiv",
    doi = "10.1103/PhysRevD.62.107504",
    journal = "Phys. Rev. D",
    volume = "62",
    pages = "107504",
    year = "2000"
}

@article{Tominaga:1999iy,
    author = "Tominaga, Kazuhiro and Saijo, Motoyuki and Maeda, Kei-ichi",
    title = "{Gravitational waves from a test particle scattered by a neutron star: Axial mode case}",
    eprint = "gr-qc/9901040",
    archivePrefix = "arXiv",
    reportNumber = "WU-AP-76-99",
    doi = "10.1103/PhysRevD.60.024004",
    journal = "Phys. Rev. D",
    volume = "60",
    pages = "024004",
    year = "1999"
}

@article{Finn:1992wt,
    author = "Finn, Lee S.",
    title = "{Detection, measurement and gravitational radiation}",
    eprint = "gr-qc/9209010",
    archivePrefix = "arXiv",
    reportNumber = "PRINT-93-0128 (NORTHWESTERN)",
    doi = "10.1103/PhysRevD.46.5236",
    journal = "Phys. Rev. D",
    volume = "46",
    pages = "5236--5249",
    year = "1992"
}

@article{Cutler:1994ys,
    author = "Cutler, Curt and Flanagan, Eanna E.",
    title = "{Gravitational waves from merging compact binaries: How accurately can one extract the binary's parameters from the inspiral wave form?}",
    eprint = "gr-qc/9402014",
    archivePrefix = "arXiv",
    reportNumber = "GRP-369",
    doi = "10.1103/PhysRevD.49.2658",
    journal = "Phys. Rev. D",
    volume = "49",
    pages = "2658--2697",
    year = "1994"
}

@article{LIGOScientific:2016aoc,
    author = "Abbott, B. P. and others",
    collaboration = "LIGO Scientific, Virgo",
    title = "{Observation of Gravitational Waves from a Binary Black Hole Merger}",
    eprint = "1602.03837",
    archivePrefix = "arXiv",
    primaryClass = "gr-qc",
    reportNumber = "LIGO-P150914",
    doi = "10.1103/PhysRevLett.116.061102",
    journal = "Phys. Rev. Lett.",
    volume = "116",
    number = "6",
    pages = "061102",
    year = "2016"
}

@article{Vishveshwara:1970zz,
    author = "Vishveshwara, C. V.",
    title = "{Scattering of Gravitational Radiation by a Schwarzschild Black-hole}",
    doi = "10.1038/227936a0",
    journal = "Nature",
    volume = "227",
    pages = "936--938",
    year = "1970"
}

@article{Konoplya:2020hyk,
    author = "Konoplya, R. A. and Zhidenko, A.",
    title = "{General parametrization of black holes: The only parameters that matter}",
    eprint = "2001.06100",
    archivePrefix = "arXiv",
    primaryClass = "gr-qc",
    doi = "10.1103/PhysRevD.101.124004",
    journal = "Phys. Rev. D",
    volume = "101",
    number = "12",
    pages = "124004",
    year = "2020"
}

@article{Unruh:1980cg,
    author = "Unruh, W. G.",
    title = "{Experimental black hole evaporation}",
    doi = "10.1103/PhysRevLett.46.1351",
    journal = "Phys. Rev. Lett.",
    volume = "46",
    pages = "1351--1353",
    year = "1981"
}

@article{Steinhauer:2015saa,
    author = "Steinhauer, Jeff",
    title = "{Observation of quantum Hawking radiation and its entanglement in an analogue black hole}",
    eprint = "1510.00621",
    archivePrefix = "arXiv",
    primaryClass = "gr-qc",
    doi = "10.1038/nphys3863",
    journal = "Nature Phys.",
    volume = "12",
    pages = "959",
    year = "2016"
}

@article{Torres:2016iee,
    author = "Torres, Theo and Patrick, Sam and Coutant, Antonin and Richartz, Mauricio and Tedford, Edmund W. and Weinfurtner, Silke",
    title = "{Observation of superradiance in a vortex flow}",
    eprint = "1612.06180",
    archivePrefix = "arXiv",
    primaryClass = "gr-qc",
    doi = "10.1038/nphys4151",
    journal = "Nature Phys.",
    volume = "13",
    pages = "833--836",
    year = "2017"
}

@article{Torres:2020tzs,
    author = "Torres, Theo and Patrick, Sam and Richartz, Maur{\'\i}cio and Weinfurtner, Silke",
    title = "{Quasinormal Mode Oscillations in an Analogue Black Hole Experiment}",
    eprint = "1811.07858",
    archivePrefix = "arXiv",
    primaryClass = "gr-qc",
    doi = "10.1103/PhysRevLett.125.011301",
    journal = "Phys. Rev. Lett.",
    volume = "125",
    number = "1",
    pages = "011301",
    year = "2020"
}

@misc{smaniotto2026blackholespectroscopygiantquantum,
      title={Black-hole spectroscopy from a giant quantum vortex}, 
      author={Pietro Smaniotto and Leonardo Solidoro and Patrik Švančara and Sam Patrick and Maurício Richartz and Carlo F. Barenghi and Ruth Gregory and Silke Weinfurtner},
      year={2026},
      eprint={2502.11209},
      archivePrefix={arXiv},
      primaryClass={gr-qc},
      url={https://arxiv.org/abs/2502.11209}, 
}

@article{Rutherford:1911zz,
    author = "Rutherford, E.",
    title = "{The scattering of alpha and beta particles by matter and the structure of the atom}",
    doi = "10.1080/14786440508637080",
    journal = "Phil. Mag. Ser. 6",
    volume = "21",
    pages = "669--688",
    year = "1911"
}

@article{Mino:2008at,
    author = "Mino, Yasushi and Brink, Jeandrew",
    title = "{Gravitational Radiation from Plunging Orbits: Perturbative Study}",
    eprint = "0809.2814",
    archivePrefix = "arXiv",
    primaryClass = "gr-qc",
    doi = "10.1103/PhysRevD.78.124015",
    journal = "Phys. Rev. D",
    volume = "78",
    pages = "124015",
    year = "2008"
}

@article{Ferrari:1984ozr,
    author = "Ferrari, Valeria and Mashhoon, Bahram",
    title = "{Oscillations of a Black Hole}",
    doi = "10.1103/PhysRevLett.52.1361",
    journal = "Phys. Rev. Lett.",
    volume = "52",
    number = "16",
    pages = "1361",
    year = "1984"
}

@article{Ferrari:1984zz,
    author = "Ferrari, Valeria and Mashhoon, Bahram",
    title = "{New approach to the quasinormal modes of a black hole}",
    doi = "10.1103/PhysRevD.30.295",
    journal = "Phys. Rev. D",
    volume = "30",
    pages = "295--304",
    year = "1984"
}

@article{Kuntz:2025gdq,
    author = "Kuntz, Adrien",
    title = {{Green function of the P{\"o}schl-Teller potential}},
    eprint = "2510.17954",
    archivePrefix = "arXiv",
    primaryClass = "gr-qc",
    doi = "10.21468/SciPostPhys.20.4.120",
    journal = "SciPost Phys.",
    volume = "20",
    number = "4",
    pages = "120",
    year = "2026"
}

@article{Zimmerman:2011dx,
    author = "Zimmerman, Aaron and Chen, Yanbei",
    title = "{New Generic Ringdown Frequencies at the Birth of a Kerr Black Hole}",
    eprint = "1106.0782",
    archivePrefix = "arXiv",
    primaryClass = "gr-qc",
    doi = "10.1103/PhysRevD.84.084012",
    journal = "Phys. Rev. D",
    volume = "84",
    pages = "084012",
    year = "2011"
}

@misc{oshita2025probingdirectwavesblack,
      title={Probing Direct Waves in Black Hole Ringdowns}, 
      author={Naritaka Oshita and Sizheng Ma and Yanbei Chen and Huan Yang},
      year={2025},
      eprint={2509.09165},
      archivePrefix={arXiv},
      primaryClass={gr-qc},
      url={https://arxiv.org/abs/2509.09165}, 
}

@article{ParticleDataGroup:2024cfk,
    author = "Navas, S. and others",
    collaboration = "Particle Data Group",
    title = "{Review of particle physics}",
    doi = "10.1103/PhysRevD.110.030001",
    journal = "Phys. Rev. D",
    volume = "110",
    number = "3",
    pages = "030001",
    year = "2024"
}

@article{Solidoro:2024yxi,
    author = "Solidoro, Leonardo and Patrick, Sam and Weinfurtner, Silke and Gregory, Ruth",
    title = "{Origin of Quasinormal Modes in Semi-Open Systems}",
    eprint = "2406.11013",
    archivePrefix = "arXiv",
    primaryClass = "gr-qc",
    doi = "10.1103/7kjp-vrml",
    journal = "Phys. Rev. Lett.",
    volume = "135",
    number = "5",
    pages = "051401",
    year = "2025"
}

@article{Poschl:1933zz,
    author = "Poschl, G. and Teller, E.",
    title = "{Bemerkungen zur Quantenmechanik des anharmonischen Oszillators}",
    doi = "10.1007/BF01331132",
    journal = "Z. Phys.",
    volume = "83",
    pages = "143--151",
    year = "1933"
}

@article{Konoplya:2022tvv,
    author = "Konoplya, R. A. and Zhidenko, A.",
    title = "{Quasinormal ringing of general spherically symmetric parametrized black holes}",
    eprint = "2201.12897",
    archivePrefix = "arXiv",
    primaryClass = "gr-qc",
    doi = "10.1103/PhysRevD.105.104032",
    journal = "Phys. Rev. D",
    volume = "105",
    number = "10",
    pages = "104032",
    year = "2022"
}

@article{Baibhav:2023clw,
    author = "Baibhav, Vishal and Cheung, Mark Ho-Yeuk and Berti, Emanuele and Cardoso, Vitor and Carullo, Gregorio and Cotesta, Roberto and Del Pozzo, Walter and Duque, Francisco",
    title = "{Agnostic black hole spectroscopy: Quasinormal mode content of numerical relativity waveforms and limits of validity of linear perturbation theory}",
    eprint = "2302.03050",
    archivePrefix = "arXiv",
    primaryClass = "gr-qc",
    doi = "10.1103/PhysRevD.108.104020",
    journal = "Phys. Rev. D",
    volume = "108",
    number = "10",
    pages = "104020",
    year = "2023"
}

@article{Nee:2023osy,
    author = {Nee, Peter James and V{\"o}lkel, Sebastian H. and Pfeiffer, Harald P.},
    title = "{Role of black hole quasinormal mode overtones for ringdown analysis}",
    eprint = "2302.06634",
    archivePrefix = "arXiv",
    primaryClass = "gr-qc",
    doi = "10.1103/PhysRevD.108.044032",
    journal = "Phys. Rev. D",
    volume = "108",
    number = "4",
    pages = "044032",
    year = "2023"
}

@book{landau2013quantum,
  title={Quantum Mechanics: Non-Relativistic Theory},
  author={Landau, L. D. and Lifshitz, E. M.},
  volume={3},
  series={Course of Theoretical Physics},
  year={2013},
  publisher={Elsevier},
  address={Amsterdam},
  isbn={978-0-08-050340-0}
}

@article{Kokkotas:1999bd,
    author = "Kokkotas, Kostas D. and Schmidt, Bernd G.",
    title = "{Quasinormal modes of stars and black holes}",
    eprint = "gr-qc/9909058",
    archivePrefix = "arXiv",
    doi = "10.12942/lrr-1999-2",
    journal = "Living Rev. Rel.",
    volume = "2",
    pages = "2",
    year = "1999"
}

@book{arfken2013mathematical,
  title={Mathematical Methods for Physicists: A Comprehensive Guide},
  author={Arfken, George B. and Weber, Hans-J{\"u}rgen and Harris, Frank E.},
  edition={7th},
  year={2013},
  publisher={Elsevier Academic Press},
  address={Amsterdam},
  isbn={978-0-12-384654-9}
}

\appendix \label{appendix}

\section{Rezzolla-Zhidenko metric}\label{app_RZ}

We adopt the Rezzolla-Zhidenko parametrization~\cite{Rezzolla:2014mua}, whose continued-fraction expansion provides an efficient description of sizable deviations using only a small number of coefficients. 
The most general static and spherically symmetric line element considered in this framework is
\begin{align}
\label{metric}
{\rm d}s^2
=
-N^2(r){\rm d}t^2
+
\frac{B^2(r)}{N^2(r)}{\rm d}r^2
+
r^2{\rm d}\Omega^2.
\end{align}
The radius $r_0$ denotes the location of the horizon, defined by $N(r_0)=0$. 
We introduce the compact radial coordinate
\begin{align}
x
\equiv
1-\frac{r_0}{r},
\end{align}
such that the horizon is located at $x=0$ and spatial infinity at $x=1$. 
The metric functions are then written as
\begin{align}
N^2(x)
&=
xA(x)\,,
\\
A(x)
&=
1-\epsilon(1-x)
+
(a_0-\epsilon)(1-x)^2
+
\widetilde{A}(x)(1-x)^3\,,
\\
B(x)
&=
1+b_0(1-x)
+
\widetilde{B}(x)(1-x)^2\,,
\end{align}
where $\widetilde{A}(x)$ and $\widetilde{B}(x)$ are represented by the continued fractions
\begin{align}
\widetilde{A}(x)
&=
\frac{a_1}{
\displaystyle
1+
\frac{a_2x}{
\displaystyle
1+
\frac{a_3x}{
\displaystyle
1+\ldots
}
}
}\,,
\\
\widetilde{B}(x)
&=
\frac{b_1}{
\displaystyle
1+
\frac{b_2x}{
\displaystyle
1+
\frac{b_3x}{
\displaystyle
1+\ldots
}
}
}\,.
\end{align}
The parameters $\epsilon$, $a_0$, and $b_0$ control the asymptotic behavior of the metric, whereas the coefficients $(a_1,a_2,\ldots)$ and $(b_1,b_2,\ldots)$ encode deviations that become progressively more important toward the horizon. 

A characteristic feature of the continued-fraction construction is its hierarchical structure. 
For $i>1$, the coefficients $a_i$ and $b_i$ affect the metric only if the preceding coefficients in the corresponding continued fraction are nonzero. 
This makes the parametrization particularly well suited to systematic truncations at finite order.

The tortoise coordinate $r^*$ is defined by
\begin{align}
\label{eq:tortoise}
\frac{{\rm d}r^*}{{\rm d}r}
=
\frac{B(r)}{N^2(r)}\,.
\end{align}
The evolution of linear perturbations depends on the RZ parameters both through the effective potential and through the relation between the areal radius $r$ and the tortoise coordinate $r^*$. 
The latter is obtained by integrating Eq.~\eqref{eq:tortoise}. For a generic set of RZ parameters, this transformation must be evaluated numerically. 
At a fixed truncation order, however, the integral can in some cases be obtained analytically.

For the analysis presented here, it is sufficient to retain the continued fractions up to second order, allowing $a_2$ and $b_2$ to be nonzero while setting all higher-order coefficients to zero. 
At this order, the relation $r^*(r)$ can be obtained analytically by decomposing the integrand into partial fractions.

\section{Time-domain integration}\label{app_TD}

Solving the time evolution of some given initial data allows us to derive the full dynamics of the system. 
For that goal, we implement a staggered leapfrog algorithm, e.g., see Ref.~\cite{Nee:2023osy}, which solves the wave equation Eq.~\eqref{eq_wave} numerically via a finite difference scheme central in time and space
\begin{multline}\label{timeintegration}
\Psi_{j}^{i} = 2\Psi_{j}^{i-1}-\Psi_{j}^{i-2}+\frac{\Delta t^2}{\Delta {r^*}^2}\left(\psi^{i-1}_{j+1}-2\psi^{i-1}_{j}+\psi^{i-1}_{j-1}\right)\\-\Delta t^{2}\,\Psi_{j}^{i-1}V_{j}\,.
\end{multline}
Here we define $\Psi^{i}_{j} = \Psi(t_{i}, r^*_{j})$, $V_{j} = V_m(r^*_{j})$. 
Moreover, $\Delta t$ and $\Delta r^*$ define the temporal and spatial resolution. 
We choose a spatial grid of 10000 equally distant points in the tortoise coordinates between $r^*=-100$ and $r^*=200$, which gives us $\Delta r^*=300/10000=0.03$. 
The time resolution is defined by $\Delta t= 1/4 \Delta r^{*}$. 
As boundary conditions, we impose $\Psi=0$ at both ends of the spatial domain. 
To avoid spurious reflections from the boundaries, we restrict the signal duration to an interval shorter than the time required for reflected waves to propagate back to the observer.

In our Gedankenexperiment, initial data is modeled as an ingoing Gaussian wave packet
\begin{align}\label{initialdata}
\Psi(t,r^*)\big|_{t=t_\mathrm{start}} &= A\exp{\left[-\left(\frac{r^* - r^*_c}{\sigma}\right)^2\right]}\,,
\\
\frac{\partial}{\partial t} \Psi(t,r^*)\big|_{t=t_\mathrm{start}} &= \frac{\partial}{\partial{r^*}} \Psi(t,r^*)\big|_{t=t_\mathrm{start}}\,,
\end{align}
which in $t=t_\mathrm{start}=0$ is centered at $r^*_c=50$.

The pulse in Eq.\eqref{initialdata} evolves toward the potential barrier. 
Initially, this injected pulse is completely independent \textit{a priori} from the background geometry. 
After being released, however, the pulse starts interacting progressively with the local space-time, until it reaches the critical point where it is scattered by the radial potential in Eq.~\eqref{eq:RW} and part of its content is reflected back to the observer position. 
As input for our Bayesian analysis in Eq.~\eqref{eq:likelihood}, we use the time series $\Psi(t, r^*=r^*_\text{obs})$, where $r^*_\text{obs}$ is the location of an observer. We fix our observer position in $r^*_\text{obs}=80$.

\section{Analytical optimal Gaussian pulse width for P\"oschl-Teller potential}\label{app_PT}

The PT potential provides an exactly solvable approximation to black-hole
scattering barriers~\cite{landau2013quantum}. Its reflection coefficient is
\begin{gather}
R(\omega)=e^{2i\omega r_0^*}
\frac{\Gamma(i\epsilon)
\Gamma(\tfrac12+i\mu-i\epsilon)
\Gamma(\tfrac12-i\mu-i\epsilon)}
{\Gamma(-i\epsilon)
\Gamma(\tfrac12+i\mu)
\Gamma(\tfrac12-i\mu)}\,,
\label{eq:Rcoeff}\\[2pt]
\epsilon=\frac{\omega}{\alpha},\qquad
\mu=\sqrt{\frac{V_0}{\alpha^2}-\frac14},\qquad
\omega_R=\alpha\mu\,,
\notag
\end{gather}
where $\mu=\omega_R/\alpha\equiv Q$ is the barrier quality factor and the only
independent dimensionless combination of $(V_0,\alpha)$.

For white noise, Parseval's theorem~\cite{arfken2013mathematical} converts
Eq.~\eqref{eq:inner_product} into
$(a|b)=(\pi S)^{-1}\int_0^\infty\Re[\tilde a^*\tilde b]\,d\omega$, exactly in the
limit that the integration window contains the full response. With the
fixed-injected-SNR normalisation the incident spectrum contributes
$|\tilde h_{\rm in}|^2\propto A^2\sigma^2e^{-\sigma^2\omega^2/2}$ with
$A^2\propto\sigma^{-1}$, so that the Fisher matrix for
$\bm{\theta}^{\mathrm{PT}}=(r^*_0,V_0,\alpha)$ is
\begin{gather}
\Gamma_{ij}(\sigma)=\kappa\,\sigma\int_0^\infty e^{-\sigma^2\omega^2/2}\,
\mathcal{W}_{ij}(\omega)\,d\omega\,,
\label{eq:fisher}\\[2pt]
\mathcal{W}_{ij}\equiv\Re\big[\partial_iR^*\partial_jR\big]
=\partial_i|R|\,\partial_j|R|+|R|^2\,\partial_i\Phi\,\partial_j\Phi\,,
\notag
\end{gather}
where $\Phi\equiv\arg R$, $\kappa$ is independent of $\sigma$, and
$\Sigma=\Gamma^{-1}$. The agreement between this analytical Fisher matrix and the
time-domain result in Fig.~\ref{fig_ptcase} validates both Parseval step and
the numerical accuracy of the time-domain code.

\medskip
\emph{\bf Universal kernel and scaling law--}
Changing variables in Eq.~\eqref{eq:fisher} to $u=\omega/\omega_R$ and setting
$x=\sigma\omega_R$, the Gaussian becomes $e^{-x^2u^2/2}$ and the prefactor
$\sigma\omega_R$ combines into $x$, so that $\Gamma\propto x\,K M(x) K$ with $K$
a diagonal, $\sigma$-independent matrix and
\begin{gather}
M_{ij}(x)=\int_0^\infty e^{-x^2u^2/2}\,\mathcal{W}_{ij}(u)\,du\,,
\label{eq:factor}\\[2pt]
\Sigma_{ii}\propto x^{-1}\big[M(x)^{-1}\big]_{ii}\,.
\notag
\end{gather}
Because $x$ enters only through the Gaussian, one has $M'(x)=-xN(x)$
\emph{identically}, where $N_{ij}$ denotes the $u^2$-weighted counterpart of
$M_{ij}$. Minimising
Eq.~\eqref{eq:factor} therefore gives the exact condition
\begin{equation}\label{eq:opt}
x_{\rm opt}^2=\frac{[M^{-1}]_{ii}}{[M^{-1}NM^{-1}]_{ii}}
\equiv\frac{1}{\langle u^2\rangle_i^{\rm eff}}\,,
\end{equation}
i.e.\ $\sigma_{\rm opt}^2\langle\omega^2\rangle_i^{\rm eff}=1$: the optimal
packet width is the inverse RMS frequency of the effective response. Since $M$
depends on the barrier only through $x$ and the dimensionless $Q$, its minimiser
is a pure number $c_i$ and
\begin{equation}\label{eq:sigmaopt}
\sigma_{\rm opt}=\frac{c_i}{\sqrt{V_{\max}}}\ \propto\ \frac{1}{\sqrt{V_0}}
\end{equation}
for every parameter. This is the universal scaling observed throughout this work.
It is worth stressing that no approximation to $M$ can modify
Eq.~\eqref{eq:sigmaopt}; they can only renormalise the number $c_i$.

\medskip
\emph{\bf Eikonal evaluation--}
For high-$Q$ barriers the reflection is nearly a pure phase below the barrier
($|R|\simeq1$) and the amplitude term in $\mathcal{W}_{ij}$ is negligible. The
Stirling expansion of Eq.~\eqref{eq:Rcoeff} then gives
\begin{gather}
\Phi\simeq Q\big(f(u)+2\alpha u\,r^*_0\big)-\tfrac{\pi}{2}\,,
\label{eq:phase}\\[2pt]
f(u)=2u\ln u+(1-u)\ln(1-u)\\-(1+u)\ln(1+u)\,,
\notag
\end{gather}
so that $\hat\phi_i\equiv\partial_i\Phi$ obeys $\hat\phi_{r_0^*}\propto u$,
$\hat\phi_{V_0}\propto\ln[(1-u)/(1+u)]$ and $\hat\phi_\alpha\propto f(u)$, all
real only for $u<1$. The kernel truncates to
$M^{(0)}_{ij}=\int_0^1e^{-x^2u^2/2}\hat\phi_i\hat\phi_j\,du$, and
Eq.~\eqref{eq:opt} yields the $\ell\to\infty$ values $c_{r_0^*}=1.31$,
$c_{V_0}=1.14$, $c_\alpha=1.74$. At finite $Q$ the marginal optima approach these
from below; for the PT barrier tuned to the Regge-Wheeler $\ell=2$ potential they
already sit at $x_{\rm opt}\simeq1$ (Fig.~\ref{fig_ptcase}), so the leading
estimate $x\simeq1$ is accurate for the quadrupole.

Two approximations separate $M^{(0)}$ from $M$: the weight $|R|^2$ is replaced by
a step, and $\partial_i|R|\partial_j|R|$ is dropped. Both are controlled. For the
first, Eq.~\eqref{eq:Rcoeff} gives $|R|^2=n_F(u)[1+\mathcal{O}(e^{-2\pi Q})]$
where $n_F=[1+e^{2\pi Q(u-1)}]^{-1}$ is a sigmoid of
fractional width $1/Q$ whose zero-width limit is the step; since $n_F-\Theta(1-u)$ is odd about $u=1$, a
Sommerfeld expansion gives a correction of \emph{second} order in $1/Q$ wherever
the response is smooth at the edge. For the second, $|R|=n_F^{1/2}$ gives
$\partial_\epsilon|R|=-\partial_\mu|R|=-\pi n_F^{1/2}(1-n_F)$, peaked at the edge
with width $1/2\pi Q$, and contributing at \emph{first} order in $1/Q$. A similar
analysis for the phase derivatives around $u=1$ produces extra terms decaying
also with $1/Q$ --- the eikonal $\hat\phi_{V_0}$ and $\hat\phi_\alpha$ diverge
logarithmically there, whereas the exact ones involve $\Re\,\psi(\tfrac12+iy)$
with $y=Q(1-u)$, which saturates at $\psi(\tfrac12)$ --- of opposite sign, so
that the two cancel. Exact quadrature of Eq.~\eqref{eq:fisher} gives at
$Q=2.5$ the conditional optima $1.330$, $1.194$, $1.687$ for $r_0^*$, $V_0$,
$\alpha$, within $3\%$, $1\%$ and $4\%$ of the eikonal values, and marginal
optima $0.99$, $0.95$, $1.22$, close to unity as displayed in
Fig.~\ref{fig_ptcase}. The eikonal kernel is therefore quantitatively adequate
already at the quadrupole, while Eq.~\eqref{eq:sigmaopt} holds at every $Q$.

\medskip
\emph{\bf Scaling law--}
Near the optimum the marginal covariance is dominated by the near-degenerate
resonant direction, sharply peaked at the band edge $u=1$ ($\omega=\omega_R$),
i.e.\ the response is approximately described by the fundamental quasinormal mode
alone. For such a peaked response the Gaussian factors out at $u=1$,
$[M(x)^{-1}]_{ii}\simeq e^{x^2/2}[M(0)^{-1}]_{ii}$, and Eq.~\eqref{eq:factor}
reduces to
\begin{gather}
\Sigma_{ii}(\sigma)\simeq C_i\,\frac{1}{\sigma}\,
\exp\Big(\tfrac12\sigma^2\omega_R^2\Big),
\label{eq:law}\\[2pt]
C_i=\big[M(0)^{-1}\big]_{ii}\big/\big(\kappa K_i^2\omega_R\big)\,.
\notag
\end{gather}
This scaling is seen for the RZ parameters in Fig.~\ref{scaling}: the derivation
therefore \emph{explains} the phenomenological form. Its two competing pieces are
physical: $\sigma^{-1}$ is the growing injected energy at fixed SNR (favouring
broad probes), while $\exp(\sigma^2\omega_R^2/2)$ is the exponential loss of
probe power at $\omega_R$ as the packet broadens (favouring narrow probes).
Minimising Eq.~\eqref{eq:law} at constant $C_i$ returns $x=1$; restoring the
residual $\sigma$-dependence shifts it to the exact value: 
\begin{equation}\label{eq:corr}
x_{\rm opt}^2=1-\frac{d\ln C_i}{d\ln x}
,
\end{equation}
The departure
of $x_{\rm opt}$ from unity thus measures how far the effective spectral weight
deviates from the fundamental quasinormal frequency.

\medskip
\emph{\bf Closed form of conditional optima--}
The \emph{conditional} uncertainties, each parameter estimated with the
others held fixed, are more tractable, since they involve only
$\Gamma_{ii}$ rather than the full inverse. Minimising
$\Sigma^{\rm cond}_{ii}=1/\Gamma_{ii}\propto[xM_{ii}]^{-1}$ gives the single-parameter
condition $x_{\rm opt,i}^2=M_{ii}/N_{ii}$. This yields
$x_{\rm opt}=1.37$, $1.20$ and $1.76$ respectively, slightly above the marginal
plateaus, the degeneracy that pulls the marginal minima toward unity being
absent here. The barrier \emph{location} closes completely:
from Eqs.~\eqref{eq:phase} and~\eqref{eq:factor}, with $a=x^2/2$,
\begin{equation}\label{eq:Mrr}
M_{rr}=\frac{\sqrt\pi\,\mathrm{erf}(\sqrt a)}{4a^{3/2}}-\frac{e^{-a}}{2a}\,,
\qquad
N_{rr}=\frac{3M_{rr}-e^{-a}}{2a}\,,
\end{equation}
so the condition reduces to $2M_{rr}=e^{-a}$, i.e.\ the transcendental
\begin{equation}\label{eq:condloc}
\sqrt{2\pi}\,\mathrm{erf}(x/\sqrt2)=(x^3+2x)e^{-x^2/2},
\quad x_{\rm opt}^{\rm cond}=1.369\,.
\end{equation}
This complementary result shows that for conditional uncertainty, i.e.\ the
regime where parameters are probed individually rather than simultaneously
marginalized, the optimal Gaussian packet width also exists and obeys a similar
relation with the barrier peak.
\end{document}